\documentclass[aps,prl,groupedaddress, twocolumn]{revtex4-1}
\usepackage{graphicx}% Include figure files
\usepackage{dcolumn}% Align table columns on decimal point
\usepackage{bm}% bold math
\usepackage{amsmath}
\begin{document}

%\preprint{}

\title{Highly Stretchable MoS$_2$ Kirigami}


\author{Paul~Z.~Hanakata}
\email[]{hanakata@bu.edu}
%\thanks{}
\affiliation{Department of Physics, Boston University, Boston, MA 02215, USA}

\author{Zenan~Qi}
\affiliation{Department of Mechanical Engineering, Boston University, Boston, MA 02215, USA}

\author{David~K.~Campbell}
\affiliation{Department of Physics, Boston University, Boston, MA 02215, USA}

\author{Harold~S.~Park}
\affiliation{Department of Mechanical Engineering, Boston University, Boston, MA 02215, USA}
\email[]{parkhs@bu.edu}

\date{\today}

\begin{abstract}
  We report the results of classical molecular dynamics simulations
  focused on studying the mechanical properties of MoS$_{2}$ kirigami.
  Several different kirigami structures were studied based upon two
  simple non-dimensional parameters, which are related to the density
  of cuts, as well as the ratio of the overlapping cut length to the
  nanoribbon length.  Our key finding is significant enhancements in
  tensile yield (by a factor of four) and fracture strains (by a
  factor of six) as compared to pristine MoS$_{2}$ nanoribbons. These
  results in conjunction with recent results on graphene suggest that
  the kirigami approach may be a generally useful one for enhancing
  the ductility of two-dimensional nanomaterials.
\end{abstract}

\pacs{}
%\keywords{}

\maketitle

Molybdenum disulfide (MoS$_{2}$) has been intensely studied in recent
years as an alternative two-dimensional (2D) material to graphene.
This interest has arisen in large part because (i) MoS$_2$ exhibits a
direct band gap of nearly 2 eV in monolayer form which is suitable for
photovoltaics~\cite{makPRL2010}; and (ii) it has recently been
explored for many potential applications, ranging from energy storage
to valleytonics
~\cite{chhowallaNC2013,wangNNANO2012,johariACSNANO2012, zengNN2012}.

The mechanical properties of MoS$_{2}$ have also been explored
recently, through both
experimental~\cite{bertolazziACSNANO2011,gomezAM2012,cooperPRB2013}
and theoretical
methods~\cite{jiangJAP2013,jiangNANO2013,dangSM2014,zhaoNANO2014}.
That MoS$_{2}$ has been reported experimentally to be more ductile
than graphene~\cite{cooperPRB2013} naturally raises the critical issue
of developing new approaches to enhancing the ductility of 2D
materials.

One approach that has recently been proposed towards this end is in
utilizing concepts of kirigami, the Japanese technique of paper
cutting, in which cutting is used to change the morphology of a
structure. This approach has traditionally been applied to bulk
materials and recently to micro-scale materials~\cite{guo2014highly,
  shyuNM2015, guoPNAS2015}, though recent
experimental~\cite{bleesN2015} and theoretical~\cite{qiPRB2014a} works
have shown the benefits of kirigami for the stretchability of
graphene.

Our objective in the present work is to build upon previous successes
in applying kirigami concepts to graphene~\cite{qiPRB2014a} to
investigate their effectiveness in enhancing the ductility of a
different 2D material, MoS$_{2}$, which is structurally more complex
than monolayer graphene due to its three-layer structure involving
multiple atom types.  We accomplish this using classical molecular
dynamics (MD) with a recently developed Stillinger-Weber
potential~\cite{jiangNANO2015}.  We find that kirigami can
substantially enhance the yield and fracture strains of monolayer
MoS$_{2}$, with increases that exceed those previously seen in
monolayer graphene~\cite{qiPRB2014a}.

%\section{Numerical Results}

%-------------FIGURE 1--------------------------------------------------------
\begin{figure}
  \centering
  \includegraphics[scale=0.4]{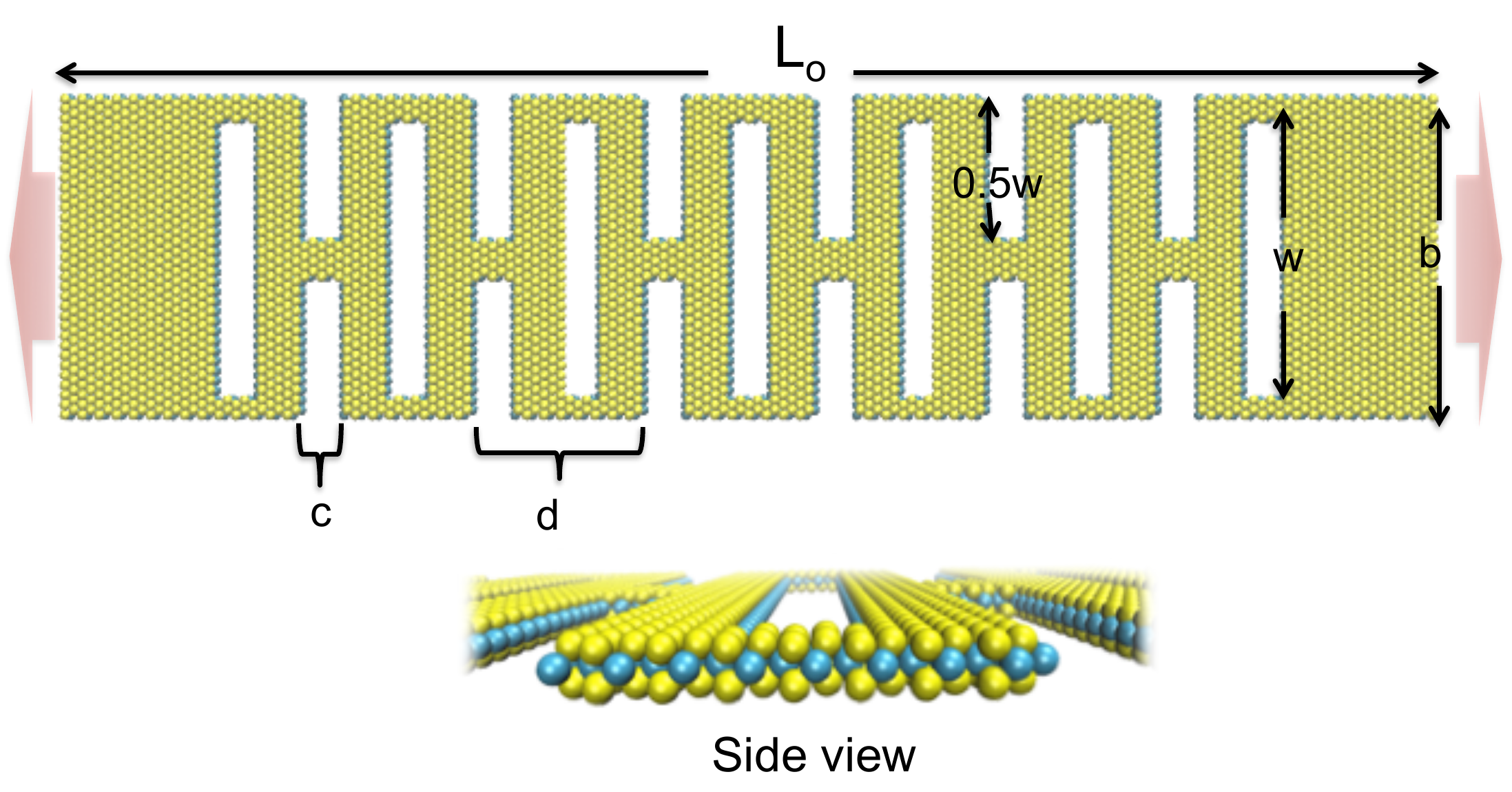}
  \caption{(Color online) Schematic of the MoS$_2$ kirigami, with key
    geometric parameters labeled. The kirigami is deformed via tensile
    displacement loading that is applied at the two ends in the
    direction indicated by the arrows.  Top image represents a top
    view of the kirigami.}
  \label{fig:schematic}
\end{figure}
%---------------------------------------------------------------------

We performed MD simulations using the Sandia-developed open source
code LAMMPS~\cite{plimptonLAMMPS,plimptonJCP1995} using the
Stillinger-Weber potential for MoS$_{2}$ of
Jiang~\cite{jiangNANO2015}.  All simulations were performed on
single-layer MoS$_{2}$ sheets.  Of relevance to the results in this
work, we note that while the Stillinger-Weber potential does not have
a term explicitly devoted to rotations, it does contain two and
three-body terms including angular dependencies, which is important
for out-of-plane deformations.  Furthermore, the Stillinger-Weber
potential of Jiang~\cite{jiangNANO2015} was fit to the phonon spectrum
of single-layer MoS$_{2}$, which includes both in and out-of-plane
vibrational motions.  As a result, the Stillinger-Weber potential
should do a reasonable job of capturing out-of-plane deformations that
involve angle changes, such as rotations.

The MoS$_{2}$ kirigami was made by cutting an MoS$_{2}$ nanoribbon,
which had free edges without additional surface treatment or
termination. A schematic view of the kirigami structure and the
relevant geometric parameters is shown in
Fig.~\ref{fig:schematic}. The key geometric parameters are the
nanoribbon length $L_{0}$, the width $b$, the height of each interior
cut $w$, the width of each interior cut $c$, and the distance between
successive cuts $d$. We considered kirigami for both zig-zag (ZZ) and
armchair (AC) edges. A representative AC MoS$_2$ kirigami consisting a
number of $N\sim 12,000$ atoms with a nanoribbon length
$L_{0}\sim 450$~\AA, width $b\sim100$~\AA, height of each interior cut
$w\sim 70$~\AA, width of each interior cut $c\sim 11$~\AA, and distance
between successive cuts $d\sim55$~\AA~is shown in
Fig.~\ref{fig:schematic}.

%---------------FIGURE 2------------------------------------------------------
\begin{figure}
  \centering
  \includegraphics[scale=0.4]{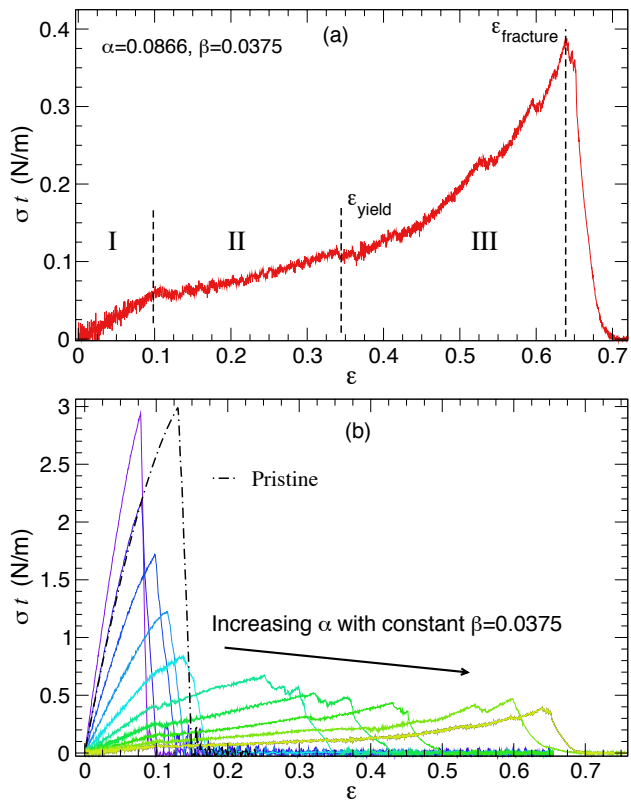}
  \caption{(Color online) Stress-strain curves of AC MoS$_{2}$
      kirigami, where the 2D stress was calculated as the stress
      $\sigma$ times simulation box $t$.  (a) Stress-strain curve for
      constant $\alpha=0.0866$, $\beta=0.0375$.  (b) Stress-strain curve
      for AC kirigami keeping $\beta=0.0375$ constant and varying
      $\alpha$.  Note the brittle fracture of the pristine MoS$_{2}$
      nanoribbon.  In general, the strain in region III increases
      substantially for $\alpha>0$.}
  \label{fig:stress-strain}
\end{figure}
%---------------------------------------------------------------------

The MD simulations were performed as follows. The kirigami was first
relaxed for 200 ps within the NVT (constant number of atoms $N$,
volume $V$ and temperature $T$) ensemble at low temperature (4.2 K),
while non-periodic boundary conditions were used in all three
directions. The kirigami was subsequently deformed in tension by
applying uniform displacement loading on both ends, such that the
kirigami was pulled apart until fracture occurred.  We note that
  in actual applications, the MoS$_{2}$ kirigami will likely lie on a
  substrate, and thus adhesive interactions with the substrate may
  impact the deformation characteristics.  In the present work, we
  focus on the intrinsic stretchability of the MoS$_{2}$ kirigami
  while leaving the interactions with a substrate for future work.

\begin{table}[h]
\small
  \caption{Comparison of mechanical properties of MoS$_2$ sheets and pristine nanoribbons in the armchair (AC) and zigzag (ZZ) direction.}
\centering
%\begin{tabular*}{| p{2cm} | p{2cm} | p{2cm} | p{2cm} |}
\begin{tabular*}{0.5\textwidth}{@{\extracolsep{\fill}}llll}
\hline 
System & $\epsilon{\rm_f}$ & $\sigma{\rm^{3D}_f (GPa)}$& $Y^{\rm 3D} ({\rm GPa})$ \\
%\\ \\ [0.5ex]
\hline 
Sheet   (AC)    &0.178   &   16.8 &      154.0 \\
Sheet   (ZZ)    &0.175 &      15.6   &    150.7\\
NR      (AC)    &0.130   &       14.6    & 145.8\\
NR      (ZZ)    &0.129  &        13.6    & 130.0 \\
%\\ [1ex] 
\hline
\end{tabular*}
\label{table:comparison}
\end{table}
%\begin{table}[h]
%\small
%  \caption{\ An example of a caption to accompany a table}
%  \label{tbl:example}
%  \begin{tabular*}{0.5\textwidth}{@{\extracolsep{\fill}}lll}
%    \hline
%    Header one/units & Header two & Header three \\
%    \hline
%    1 & 2 & 3 \\
%    4 & 5 & 6 \\
%    7 & 8 & 9 \\
%    10 & 11 & 12 \\
%    \hline
%  \end{tabular*}
%\end{table}
In addition, we simulated MoS$_2$ sheets (defined as monolayer
  MoS$_{2}$ with periodic boundary conditions in the plane) and
pristine nanoribbons with no cuts for comparative purposes. The
calculated fracture strains $\epsilon{\rm_f}$, fracture stresses
$\sigma{\rm^{3D}_f}$, and Young's modulus $Y^{\rm 3D}$ are tabulated
in Table~\ref{table:comparison}. The results are in reasonably good
agreement with the experimental and first-principles studies of
MoS$_2$ monolayer sheets~\cite{bertolazziACSNANO2011,
  cooperPRB2013}.\footnote[4]{In the previous table, 3D stresses
  $\sigma{\rm^{3D}_f}$ are calculated as
  $\sigma{\rm^{2D}_f}/t_{\rm h}$, where $t_{\rm h}$ is the effective thickness
  with a value of $\sim6$~\AA.}

In Figure~\ref{fig:stress-strain} (a), we plot a representative
stress-strain curve of MoS$_2$ kirigami.  For this, and the subsequent
discussion, we introduce two non-dimensional geometric parameters
$\alpha=(w-0.5b)/L_{0}$ and $\beta=(0.5d-c)/L_{0}$, which were also
previously used to describe graphene
kirigami~\cite{qiPRB2014a}. $\alpha$ represents the ratio of the
overlapping cut length to the nanoribbon length, while $\beta$
represents the ratio of overlapping width to the nanoribbon length.
Put another way, $\alpha$ describes the geometry orthogonal to the
loading direction, while $\beta$ describes the geometry parallel to
the loading direction.  Figure~\ref{fig:stress-strain}(a) shows the
stress-strain for the specific choices of $\alpha=0.0866$, and
$\beta=0.0375$, which were obtained by choosing $b$=101.312~\AA ,
$L_{0}$=438.693~\AA, $w$=88.648~\AA, $c$=10.967~\AA, and
$d$=54.837~\AA.  In contrast, Figure~\ref{fig:stress-strain}(b) shows
the change in the stress-strain response if $\beta=0.0375$ is kept
constant while $\alpha$ changes.  This is achieved by changing $w$
while keeping other geometric parameters constant.  We also note that
the 2D stress was calculated as stress times simulation box size
perpendicular to the plane $\sigma\times t$ to remove any issues in
calculating the thickness~\cite{jiangNANO2013}, where the stress was
obtained using the virial theorem, as is done in LAMMPS.

\begin{figure}
  \centering
  \includegraphics[scale=0.37]{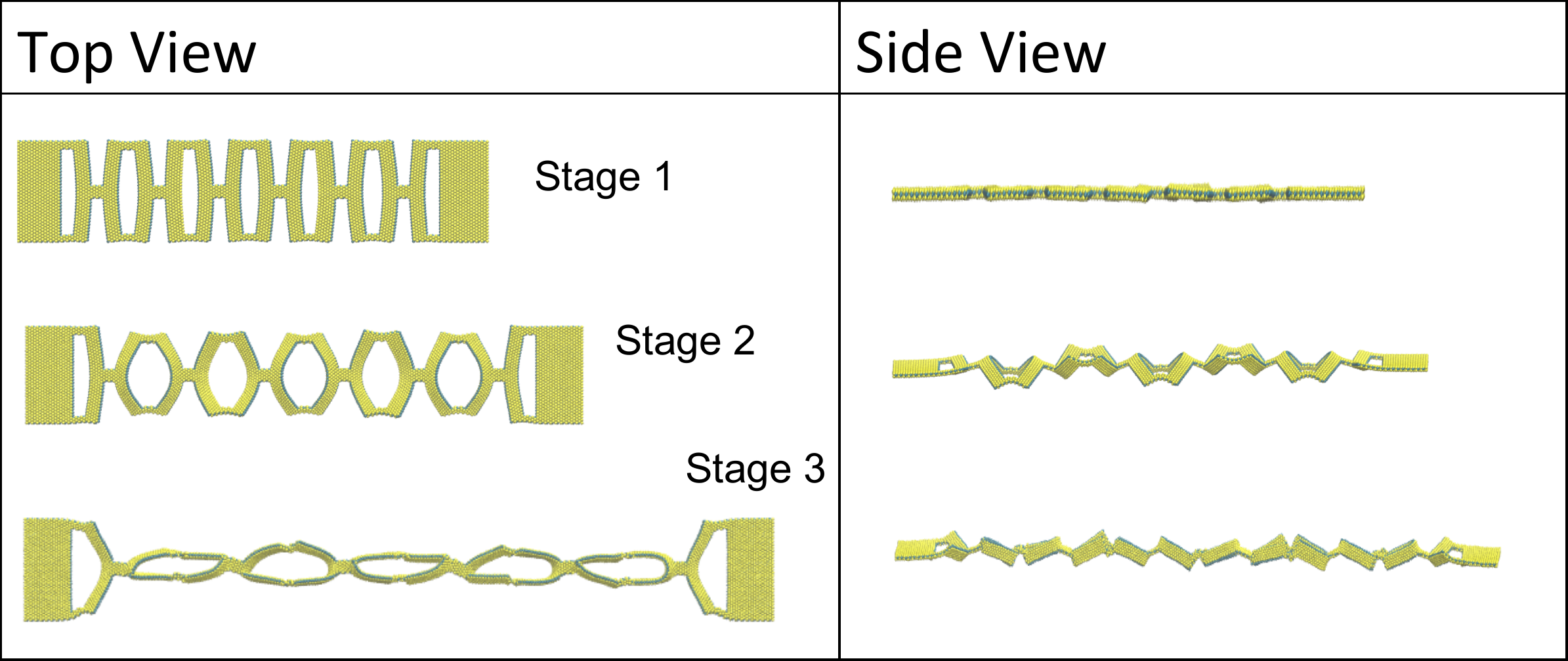}
  \caption{Side and top views of kirigami during deformation. }
  \label{fig:stages}
\end{figure}

It can be seen that there are generally three major stages of
deformation for the kirigami, as separated by the dashed lines in
Fig. \ref{fig:stress-strain}(a). In the first stage (region I), the
deformation occurs via elastic bond stretching, and neither flipping
nor rotation of the monolayer MoS$_{2}$ sheet is observed as shown in
Fig.~\ref{fig:stages}. In previous work, it was found that graphene
kirigami rotates and flips in the first stage instead of bond
stretching~\cite{qiPRB2014a}.  This does not occur for kirigami in
MoS$_2$ in this first stage because the bending modulus of MoS$_2$ is
nearly seven times higher than that of graphene~\cite{jiangNANO2013}.

%------------------FIGURE 3---------------------------------------------------
%---------------------------------------------------------------------

In the second stage (region II), for tensile strains ($\epsilon$)
exceeding about 10$\%$, further strain hardening occurs.  Kirigami
patterning allows the MoS$_{2}$ monolayer to exhibit out-of-plane
deflections, as shown in Fig.~\ref{fig:stages}, which allows the
MoS$_{2}$ monolayer to undergo additional tensile deformation, which
is in contrast to the brittle fracture observed for the pristine
nanoribbon immediately following the initial yielding event as shown
in Fig.~\ref{fig:stress-strain}(b).  Furthermore, the out-of-plane
deflections cause the slope of the stress-strain curve in region II to
be smaller than that in region I.  This is because of the change in
deformation mechanism from purely elastic stretching of bonds in
region I, to a combination of stretching and out of plane buckling in
region II.

%------------------FIGURE 3---------------------------------------------------
\begin{figure}
  \centering
  \includegraphics[scale=0.28]{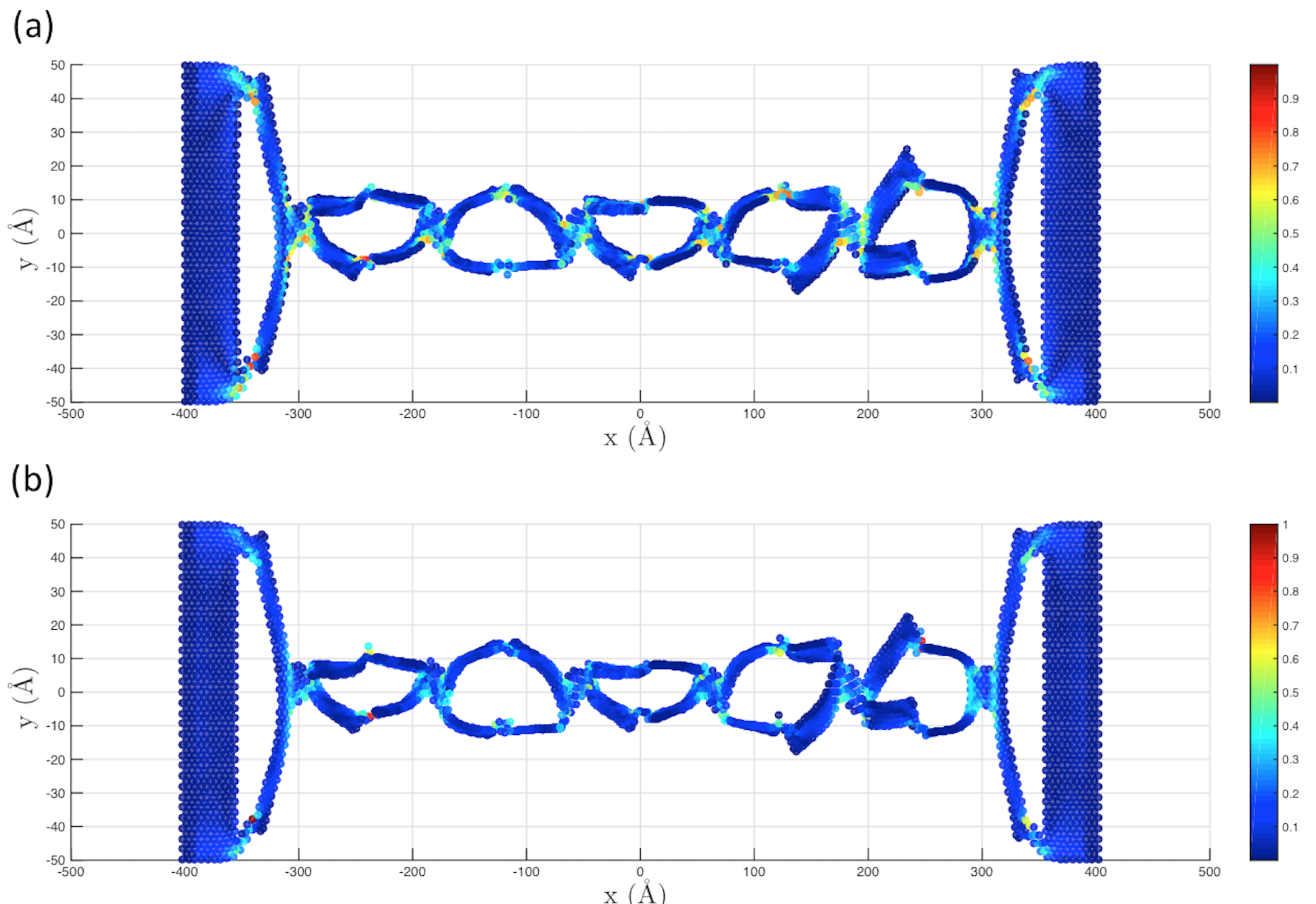}
  \caption{Von Mises stress prior to the fracture at a tensile strain
    of 62\% in (a) Mo layer and (b) top S layer of kirigami in
    Fig.~\ref{fig:stages}. We plot the stress distribution layer by
    layer to give a clear picture of the stress distribution. The von
    Mises stress were scaled between 0 and 1.  }
  \label{fig:vM}
\end{figure}
%---------------------------------------------------------------------

Local bond breaking near the edges starts to occur at the tensile
strain of $\epsilon=35\%$. The occurrence of bond breaking is usually
defined as the yield point, and signifies the demarkation between
regions II and III. This local bond breaking occurs due to the
concentrated stress at the edges connecting each slab, as previously
observed in graphene kirigami~\cite{qiPRB2014a}.  At this stage, each
kirigami unit is held by a small connecting ribbon which allows the
monolayer to be almost foldable. Fig.~\ref{fig:stages} (stages 1 to 3)
shows how the inner cut surface area having initial area $w\times c$
and the height of the monolayer (largest out-of-plane distance between
S atoms) can change significantly during the tensile elongation.

In the final stage, after more than $62.5\%$ tensile strain, fracture
and thus failure of the kirigami nanoribbon is observed.  Unlike the
pristine nanoribbon, the yield point can differ substantially from the
fracture strain and the difference increases with increasing
cut-overlap, which was described previously as shown in
Fig.~\ref{fig:stress-strain}(b). Thus, it is important to quantify the
yield point of the kirigami as it defines the beginning of the
irreversible deformation regime.  Note that these regions vary
depending on the kirigami structure as shown in
Fig.~\ref{fig:stress-strain}(b).  

We also show, in Fig.~\ref{fig:vM}, the von Mises stress distribution
prior to fracture at a tensile strain of 62\%.  In Fig.~\ref{fig:vM},
the stress values were scaled between 0 and 1, and the stress
distributions in the top S layer and single Mo layer were plotted
separately for ease of viewing as MoS$_{2}$ has a tri-layer structure.
We found that the largest stresses are concentrated near the edges of
the each kirigami unit cell similar to that previously observed in
graphene kirigami~\cite{qiPRB2014a}.

%---------------------------------------------------------------------
\begin{figure}
  \centering
  \includegraphics[scale=0.4]{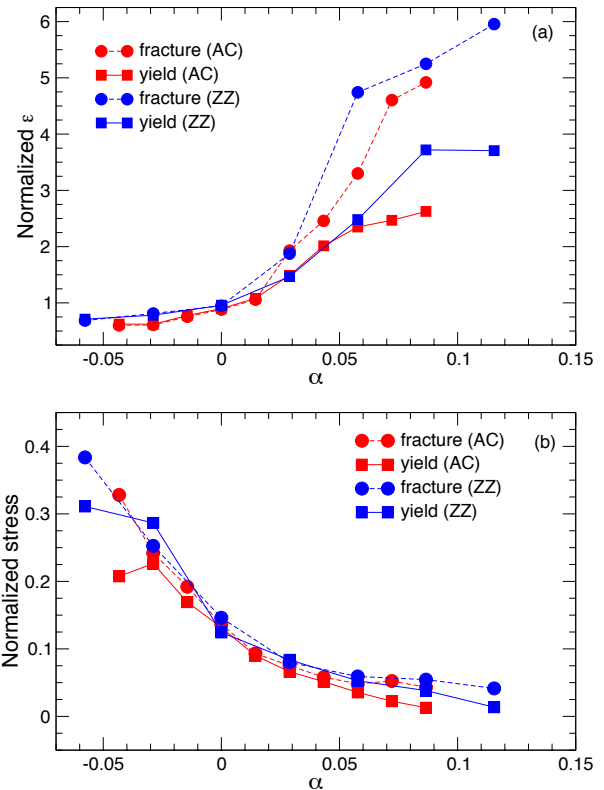}
  \caption{(Color online) (a) Influence of $\alpha$ on yield and
    fracture strain for zigzag (ZZ) and armchair (AC) MoS$_2$
    kirigami, with constant $\beta=0.0375$ for AC and constant
    $\beta=0.0417$ for ZZ.  (b) Influence of $\alpha$ on yield and
    fracture stress for zigzag (ZZ) and armchair (AC) MoS$_2$
    kirigami. Data are normalized by MoS$_{2}$ nanoribbon results
    with the same width.}
  \label{fig:alpha}
\end{figure}
%---------------------------------------------------------------------

Having established the general deformation characteristics for
MoS$_{2}$ kirigami, we now discuss how the yield and failure
characteristics are dependent on the specific kirigami geometry.  We
discuss the yield and fracture stresses and strains in terms of the
two geometric parameters $\alpha$ and $\beta$ that were previously
defined.

The yield strain as a function of $\alpha$ is shown in Fig.
\ref{fig:alpha}(a), while the yield stress as a function of $\alpha$
is shown in Fig. \ref{fig:alpha}(b). In these, and all subsequent
figures, the stresses and strains are normalized by those for pristine
MoS$_{2}$ nanoribbons of the same width such that the effect of the
kirigami parameters can be directly quantified. As shown in Fig.
\ref{fig:alpha}, the MoS$_{2}$ kirigami becomes significantly more
ductile for $\alpha>0$, where the zigzag chirality reaches a yield
strain that is about a factor of 6 larger than the pristine
nanoribbon. In contrast, Fig. \ref{fig:alpha} (b) shows that the yield
stress for kirigami correspondingly decreases dramatically for
increasing $\alpha$.  We also note that the kirigami patterning
appears to have a similar effect on the ductility of zigzag and
armchair MoS$_{2}$ kirigami (shown in Fig. \ref{fig:alpha}(a)) as the
fracture strain and bending modulus of MoS$_2$ monolayer sheet in
zigzag and armchair direction are similar~\cite{jiangNANO2015,
  jiangNANO2013}.

%---------------------------------------------------------------------
\begin{figure}
  \centering
  \includegraphics[scale=0.4]{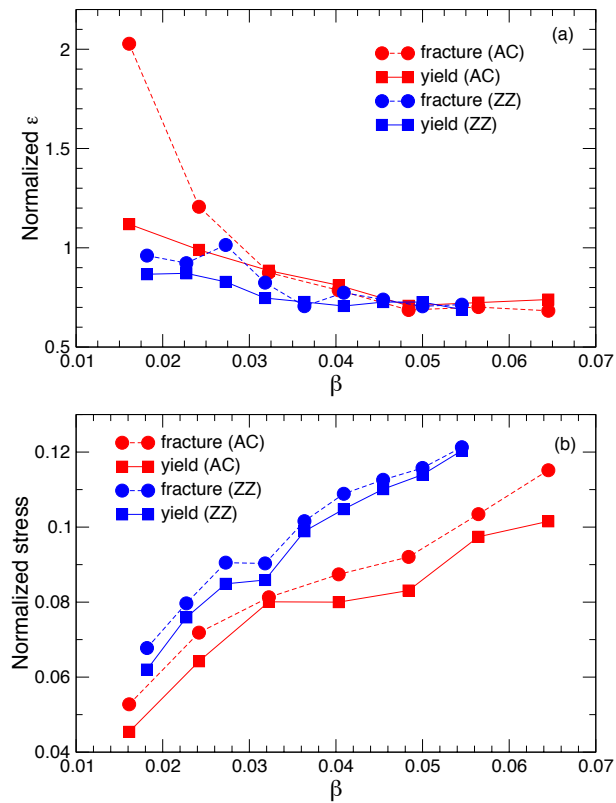}
  \caption{(Color online) Influence of $\beta$ on the kirigami yield
    and fracture strain (a) and stress (b), with constant
    $\alpha=0.0186$ for AC and constant $\alpha=0.0157$ for ZZ. Data
    are normalized by MoS$_2$ nanoribbon results with the same width.}
  \label{fig:beta}
\end{figure}
%---------------------------------------------------------------------
  
The increased ductility occurs because $\alpha=0$ corresponds
to the case when the edge and interior cuts begin to overlap.
Increasing $\alpha$ above zero corresponds to when the edge and
interior cuts do overlap, and thus it is clear that increasing the
overlap increases the ductility of the MoS$_{2}$ kirigami. In
contrast, the yield stress is higher for smaller $\alpha$ because for
negative $\alpha$, the edge and interior cuts do not overlap, and thus
the deformation of the kirigami more closely resembles that of the
cut-free nanoribbon.

In addition to the results of $\alpha$, the effect of $\beta$ on the
kirigami ductility is shown in Figs. \ref{fig:beta}(a) and
\ref{fig:beta}(b). Specifically, $\beta$ is varied by changing $d$
while keeping other geometric parameters constant. For both the yield
stress and strain, $\beta$ does impact the yield stress and strain.
Increasing $\beta$ corresponds to an increase in the overlapping
region width, which thus results in a smaller yield strain, and
increased yield stress as compared to a pristine nanoribbon. For
$\beta\geq0.03$, we do not observe large differences between the AC
and ZZ behavior in the case of varying $\beta$ because increasing
$\beta$ (or decreasing the cut density) makes the kirigami more
pristine, leading to similar values of fracture stress and strain in
the AC or ZZ direction (see Table~\ref{table:comparison}).  Our
results suggest that the failure strain can be maximized by increasing
the overlapping cut (increasing $\alpha$) and increasing density of
the cuts (decreasing $\beta$).

%  \dkc{Did Guo et al actually introduce cuts or simply have metallic "trellises"/networks
%  that had different aspect ratios? I'm on a plane and don't have their paper in front of me,
%  but I recall they did do this. If so, we need to rephrase the ensuing paragraph slightly. The 
%  paragraph is very important and well written.}
Recently, Guo et al. showed stretchability of metal electrodes can be
enhanced by creating geometries similar to the ones illustrated in
Fig.~\ref{fig:schematic}~\cite{guoPNAS2015}.  Adopting the geometric
ratios determining fracture strain described in
Ref.~\cite{guoPNAS2015}, we found similar trends: the fracture strain
increases with decreasing $\frac{(b-w)}{c}$ and increases with
increasing $\frac{b}{d}$. It is interesting to see that a similar
trend is operant at a different length scale (an atomically-thin
monolayer in this work as compared to a $\approx$40 nm thin film in
the work of Guo et al.), and for a different material system
(MoS$_{2}$ in this work, nanocrystalline gold in the work of Guo et
al.), which suggests that the fracture strain in patterned membranes
can be described entirely by geometric parameters.

It is also interesting to note that the yield and fracture strain
enhancements shown in Fig. \ref{fig:alpha}(a) exceed those previously
reported for monolayer graphene kirigami~\cite{qiPRB2014a}.  The main
reason for this is that the failure strain for the normalizing
constant, that of a pristine nanoribbon of the same width, is smaller
for MoS$_{2}$.  As shown in Table \ref{table:comparison}, this value
is about 13\%, whereas the value for a pristine graphene nanoribbon
was found to be closer to 30\%~\cite{qiPRB2014a}.  However, the
largest failure strain for the MoS$_{2}$ and graphene kirigami were
found to be around 65\%, so the overall failure strains for graphene
and MoS$_{2}$ kirigami appear to reach similar values.

\begin{figure}
  \centering
  \includegraphics[scale=0.3]{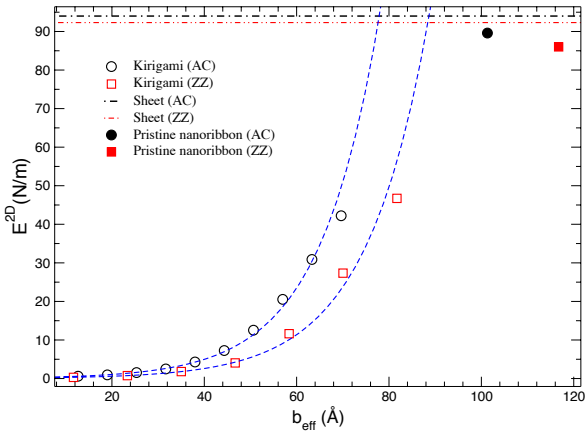}
  \caption{2D Young's modulus $E^{\rm 2D}$ of armchair (AC) and zigzag
    (ZZ) kirigami, pristine nanoribbons (PNR), and sheets. Inset shows
    $E^{\rm 2D}$ of kirigami normalized by PNRs. The fitting dashed
    line (colored blue) is given as a guide to the eye. }
  \label{fig:YM}
\end{figure}

In addition to the yield and fracture behavior, we also discuss the
elastic properties, or Young's modulus.  For the kirigami system, we
expect the Young's modulus to decrease with increasing width of the
cut $w$ due to edge effects~\cite{jiangJAP2013}. Fig.~\ref{fig:YM}
shows the dependence of Young's modulus with effective width
$b_{\rm eff}=b-w$. As can be seen for both armchair and zigzag
orientations, the modulus decreases nonlinearly with decreasing
effective width, reaching a value that is nearly 200 times smaller than
the corresponding bulk value for the smallest effective width value we
examined. Furthermore, the trend of the decrease differs from that
previously seen in graphene nanoribbons based on first principles
calculations~\cite{wagnerPRB2011} and MoS$_2$ nanoribbons based on
atomistic simulations~\cite{jiangJAP2013}, where a significantly more
gradual decrease in stiffness was observed.  This is due to the fact
that for a given nanoribbon width $b$, the kirigami has significantly
more edge area than a nanoribbon, leading to significant decreases in
elastic stiffness even for effective widths $b_{\rm eff}$ that are
close to the corresponding nanoribbon width.

%---------------------------------------------------------------------
%---------------------------------------------------------------------

\begin{figure}
  \centering
  \includegraphics[scale=0.32]{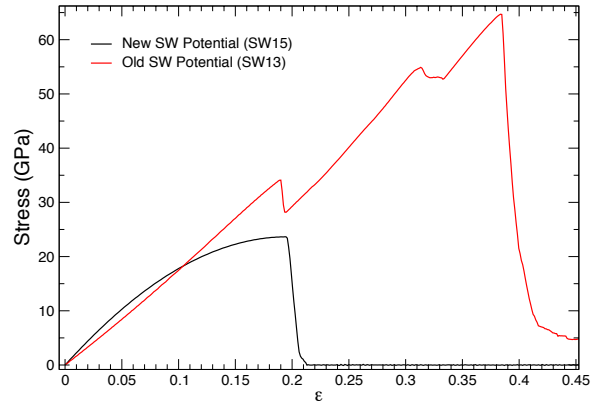}
  \caption{(Color online) Stress-strain curve of a monolayer MoS$_2$
    sheet under tensile loading along the armchair direction using two
    different SW potentials. The newer SW
    potential~\cite{jiangNANO2015} matches better with the trends
    observed in DFT calculations~\cite{cooperPRB2013} than the first
    SW potential of Jiang et al.~\cite{jiangJAP2013}.  No phase
    transition is observed with the more recent SW potential of
    Jiang~\cite{jiangNANO2015}. For SW13, breaking of bonds between the Mo and S layers occur at
      $\epsilon\sim$0.2 and $\epsilon\sim$0.3 as observed in
      Ref.~\cite{zhaoNANO2014}}
  \label{fig:SW}
\end{figure}

%\pzh{The expression that I proposed does
%  not coincide with the pristine's value (see new figure). So we might need
%  to rescale them with the pristine values. So we can write
%  $E^{\rm 2D}=a_0\exp\Big[\frac{b_{\rm eff}}{b_{\rm 0}}\Big]$}
Before concluding, we note that we have used the more
recent Stillinger-Weber (SW15) potential of Jiang~\cite{jiangNANO2015}
rather than the earlier SW potential also developed by Jiang and
co-workers~\cite{jiangJAP2013} (SW13).  This is because in comparing
the tensile stress-strain curves, the SW15 potential more closely
captured the trends observed in DFT
calculations~\cite{cooperPRB2013}. We have also performed simulations
of many kirigamis, nanoribbons, and monolayer sheets using the old SW
potential. We have found qualitatively similar results {\it with the
  very important difference} that the SW13 potential predicts a
tensile phase transition in pristine nanoribbon and monolayer
sheet~\cite{zhaoNANO2014} that is not observed in the SW15
potential~\cite{jiangNANO2015}.
%The primary difference we have found
%is that the older SW potential predicts a tensile phase
%transformation~\cite{zhaoNANO2014,dangSM2014} that is not observed in the newer
%SW potential~\cite{jiangNANO2015}.  
A comparison of the tensile stress-strain curve for monolayer
MoS$_{2}$ is shown in Fig.~\ref{fig:SW} for the potentials of Jiang
(SW15)~\cite{jiangNANO2015}, and Jiang et
al. (SW13)~\cite{jiangJAP2013}.

%---------------------------------------------------------------------
%---------------------------------------------------------------------
%\section{Conclusion}
In summary, we have applied classical molecular dynamics simulations
to demonstrate that the kirigami patterning approach can be used to
significantly enhance the tensile ductility of monolayer MoS$_{2}$,
despite the much higher bending modulus and rather more complex
tri-layer structure of MoS$_{2}$ compared to graphene.  The resulting enhancements
in tensile ductility are found to exceed those previously reported for
graphene~\cite{qiPRB2014a}.  These results suggest that kirigami may
be a broadly applicable technique for increasing the tensile ductility
of two-dimensional materials generally, and for opening up the
possibility of stretchable electronics and photovoltaics using
monolayer MoS$_{2}$.

\begin{acknowledgments}
P.Z.H. and Z.Q. acknowledge the support of the Physics and Mechanical
Engineering Departments at Boston University.  D.K.C. is grateful for
the hospitality of the Aspen Center for Physics which is supported by
NSF Grant $\#$PHY-1066293, and of the International Institute for
Physics of the Federal University of Rio Grande do Norte, in Natal,
Brazil, where some of this work was completed.  
\end{acknowledgments}

\bibliography{mos2}

\end{document}